# Heat flow in composite rods – an old problem reconsidered

T. Kranjc[a)]
*Department of Physics and Technology, Faculty of Education,*
*University of Ljubljana, Kardeljeva ploscad 16, 1000 Ljubljana, Slovenia*

and

J. Peternelj
*Faculty of Civil and Geodetic Engineering,*
*University of Ljubljana, Jamova 2, 1001 Ljubljana, Slovenia*

The interface temperature of two rods with equal cross section joined at one end and with different initial temperatures, initially always acquires the value characteristic for two semi-infinite rods. This value, which is shown to be a consequence of energy conservation is, in general, different from the thermal equilibrium temperature in finite rods. To illustrate this result, two particular cases are discussed.

## I. INTRODUCTION

In this brief paper we wish to examine certain aspects of the well known problem of linear flow of heat in composite rods. The usual treatment of the problem gives a mathematical solution but lacks a clear physical picture. It is our goal to clarify a certain point concerning the time dependence of interface temperature and the approach to thermal equilibrium which, at least to our knowledge, has not been addressed and clarified in the past. We give a clear explanation together with a clean mathematical solution that confirms it.

Consider two semi-infinite rods of equal and uniform cross section joined at one end and insulated on the sides. If the initial temperatures are $T_{01}$ and $T_{02} > T_{01}$, the corresponding temperature distribution in the rods for $t > 0$ is well known to be[1,2] (for the benefit of the reader a brief derivation is presented in the Appendix)

$$T_1(x,t) = \frac{\kappa T_{01} + T_{02}}{\kappa + 1} - \frac{T_{02} - T_{01}}{\kappa + 1} erf\left(\frac{-x}{2\sqrt{\chi_1 t}}\right), \quad x < 0, \tag{1a}$$

and

$$T_2(x,t) = \frac{\kappa T_{01} + T_{02}}{\kappa + 1} + \frac{\kappa(T_{02} - T_{01})}{\kappa + 1} erf\left(\frac{x}{2\sqrt{\chi_2 t}}\right), \quad x > 0, \tag{1b}$$

$\kappa = \frac{k_1}{k_2}\sqrt{\frac{\chi_2}{\chi_1}}$ where $k_1$ and $k_2$ are thermal conductivities of the left ($x < 0$) and the right ($x > 0$) rod, and $\chi_1$ and $\chi_2$ are the corresponding thermal diffusivities. From the above equations it follows

$$\lim_{t \to \infty} T_1(x,t) = \lim_{t \to \infty} T_2(x,t) = T_1(0, t > 0) = T_2(0, t > 0) = \frac{\kappa T_{01} + T_{02}}{\kappa + 1} \tag{2}$$

which is saying that the interface temperature is constant and equal to the final equilibrium temperature of the rods.

On the other hand, if the rods are of finite and equal length $L$, the equilibrium temperature, obtained by an elementary calculation using the conservation of energy is

$$T_{eq} = \frac{\rho_1 c_1 T_{01} + \rho_2 c_2 T_{02}}{\rho_1 c_1 + \rho_2 c_2} = \frac{\left(\sqrt{\chi_2/\chi_1}\right)\kappa T_{01} + T_{02}}{\left(\sqrt{\chi_2/\chi_1}\right)\kappa + 1} \qquad (3)$$

where $\rho_i$ and $c_i$ are the density and the specific heat of the respective rod. This result which does not depend on the length of the rods seems to be at odds with the corresponding result for semi-infinite rods given by eq. (2).

## II. TEMPERATURE DISTRIBUTION IN COMPOSITE RODS OF FINITE LENGTH

Consider two rods of lengths $L_1$ and $L_2$, and of equal cross-section aligned with the $x$-axis and joined at $x = 0$. The left rod with the initial temperature $T_{01}$ covers the interval $-L_1 < x < 0$, and the right rod with the initial temperature $T_{02} > T_{01}$ covers the interval $0 < x < L_2$. Assuming that the rods are insulated on the sides implies that $T = T(x, t)$. Consequently, the temperature distribution is determined by the heat diffusion equation of the form

$$\frac{\partial T_i}{\partial t} = \chi_i \frac{\partial^2 T_i}{\partial x^2}, \quad i = 1, 2, \qquad (4)$$

together with the initial conditions

$$T_1(x, 0) = T_{01}, \quad T_2(x, 0) = T_{02}, \qquad (5a)$$

the boundary conditions at the ends,

$$\left.\frac{\partial T_1}{\partial x}\right|_{x=-L_1} = \left.\frac{\partial T_2}{\partial x}\right|_{x=L_2} = 0, \qquad (5b)$$

and the boundary conditions at the interface,

$$T_1(0, t > 0) = T_2(0, t > 0), \qquad (5c)$$

$$-k_1 \left.\frac{\partial T_1}{\partial x}\right|_{x=0} = -k_2 \left.\frac{\partial T_2}{\partial x}\right|_{x=0}. \qquad (5d)$$

Using the Laplace transformation[3],

$$\tilde{T}(x,s) = \int_0^\infty e^{-st} T(x,t) dt,$$

the heat diffusion equation in terms of the Laplace transform $\tilde{T}(x,s)$ is the following ordinary inhomogeneous second order differential equation.

$$-\chi_i \frac{d^2 \tilde{T}_i(x,s)}{dx^2} - s\tilde{T}(x,s)_i = -\tilde{T}_i(x,0), \quad i = 1, 2. \tag{6}$$

The corresponding initial and boundary conditions are

$$\tilde{T}_1(x,0) = T_{01}, \quad \tilde{T}_2(x,0) = T_{02}, \tag{7a}$$

$$\tilde{T}_1(0,s) = \tilde{T}_2(0,s), \tag{7b}$$

$$\left. \frac{d\tilde{T}_1}{dx} \right|_{x=-L_1} = \left. \frac{d\tilde{T}_2}{dx} \right|_{x=L_2} = 0, \tag{7c}$$

$$-k_1 \left. \frac{d\tilde{T}_1}{dx} \right|_{x=0} = -k_2 \left. \frac{d\tilde{T}_2}{dx} \right|_{x=0}. \tag{7d}$$

The solution of (6) satisfying the initial and boundary conditions (7) is

$$\tilde{T}_1(x,s) = \frac{T_{01}}{s} + \left(\frac{T_{02} - T_{01}}{s}\right) \cdot \left\{ \frac{\cosh\left((x+L_1)\sqrt{\frac{s}{\chi_1}}\right) \sinh\left(L_2 \sqrt{\frac{s}{\chi_2}}\right)}{\cosh\left(L_1 \sqrt{\frac{s}{\chi_1}}\right) \sinh\left(L_2 \sqrt{\frac{s}{\chi_2}}\right) + \kappa \sinh\left(L_1 \sqrt{\frac{s}{\chi_1}}\right) \cosh\left(L_2 \sqrt{\frac{s}{\chi_2}}\right)} \right\},$$
$$-L_1 < x < 0, \quad (8a)$$

$$\tilde{T}_2(x,s) = \frac{T_{02}}{s} - \kappa\left(\frac{T_{02} - T_{01}}{s}\right) \cdot \left\{ \frac{\cosh\left((x-L_2)\sqrt{\frac{s}{\chi_2}}\right) \sinh\left(L_1 \sqrt{\frac{s}{\chi_1}}\right)}{\cosh\left(L_1 \sqrt{\frac{s}{\chi_1}}\right) \sinh\left(L_2 \sqrt{\frac{s}{\chi_2}}\right) + \kappa \sinh\left(L_1 \sqrt{\frac{s}{\chi_1}}\right) \cosh\left(L_2 \sqrt{\frac{s}{\chi_2}}\right)} \right\},$$
$$0 < x < L_2. \quad (8b)$$

In particular, taking the limit $L_1, L_2 \to \infty$ we get

$$\lim_{L_1, L_2 \to \infty} \tilde{T}_1(x,s) \equiv \tilde{T}_1^{(\infty)}(x,s) = \frac{T_{01}}{s} + \frac{T_{02} - T_{01}}{s(1+\kappa)} \cdot e^{\sqrt{\frac{s}{\chi_1}} x}, \tag{9a}$$

$$\lim_{L_1, L_2 \to \infty} \tilde{T}_2(x,s) \equiv \tilde{T}_2^{(\infty)}(x,s) = \frac{T_{02}}{s} - \kappa \frac{T_{02} - T_{01}}{s(1+\kappa)} \cdot e^{-\sqrt{\frac{s}{\chi_2}} x}. \tag{9b}$$

These limits represent the Laplace transforms of (1a) and (1b) respectively.

Applying the final value theorem for Laplace transforms[3],

$$\lim_{t \to \infty} T(x,t) = \lim_{s \to 0} s\tilde{T}(x,s),$$

to the eqs. (9) we recover the result (2). When we apply this theorem to eqs. (8) we obtain

$$\lim_{t \to \infty} T_1(x,t) = \lim_{t \to \infty} T_2(x,t) \equiv T'_{eq} = \frac{\left((L_1/L_2)\sqrt{\chi_2/\chi_1}\right)\kappa T_{01} + T_{02}}{\left((L_1/L_2)\sqrt{\chi_2/\chi_1}\right)\kappa + 1}, \qquad (10)$$

which reduces to (3) for $L_1 = L_2$ and to (2) for $L_1/L_2 = \sqrt{\chi_1/\chi_2}$.

To find the inverse Laplace transforms of (8a) and (8b) we use the residue theorem[3], i.e.,

$$T(x,t) = \sum \text{residues of } e^{st}\tilde{T}(x,s),$$

where the sum runs over all poles of $\tilde{T}(x,s)$.

Let $\sigma = \dfrac{L_2}{L_1}\sqrt{\dfrac{\chi_1}{\chi_2}}$ and $i\beta = L_1\sqrt{\dfrac{s}{\chi_1}}$. From the expressions (8a) and (8b) we deduce that $\tilde{T}_1(x,s)$ and $\tilde{T}_2(x,s)$ are single-valued functions of $s$ with simple poles at $s = 0$ and $s = -\dfrac{\chi_1}{L_1^2}\beta_m^2$ where $\pm\beta_m$, $m = 1, 2, 3,...$, are the roots of the equation

$$\cos\beta \sin\sigma\beta + \kappa \sin\beta \cos\sigma\beta = 0, \qquad (11)$$

which is obtained by equating denominator of (8a) or (8b) to zero.

In what follows we will consider two cases only. Case 1 where we set $L_1 = L_2 = L$ and Case 2 where we choose $L_1/L_2 = \sqrt{\chi_1/\chi_2}$.

**Case 1**:

Applying the residue theorem in this case (noting that in this case $\sigma = \sqrt{\chi_1/\chi_2}$), it follows

$$T_1(x,t) = T_{eq} + 2(T_{02} - T_{01})\sum_{m=1}^{\infty} \frac{1}{\beta_m} \cdot \frac{e^{-\chi_1\beta_m^2 t/L^2} \sin\sigma\beta_m \cos\beta_m(1+x/L)}{[(\sigma+\kappa)\cos\beta_m \cos\sigma\beta_m - (1+\sigma\kappa)\sin\beta_m \sin\sigma\beta_m]}, \qquad (12a)$$

and

$$T_2(x,t) = T_{eq} - 2\kappa(T_{02} - T_{01})\sum_{m=1}^{\infty} \frac{1}{\beta_m} \cdot \frac{e^{-\chi_1\beta_m^2 t/L^2} \sin\beta_m \cos\sigma\beta_m(x/L-1)}{[(\sigma+\kappa)\cos\beta_m \cos\sigma\beta_m - (1+\sigma\kappa)\sin\beta_m \sin\sigma\beta_m]} \qquad (12b)$$

where $T_{eq}$ is defined by (3). Moreover, the eq. (11) can be rewritten as

$$\sin\beta \sin\sigma\beta(\cot\beta + \kappa\cot\sigma\beta) = 0. \qquad (13)$$

The roots of (11) are thus seen to be the roots of equation

$$\cot \beta + \kappa \cot \sigma \beta = 0 \tag{14a}$$

which are all real and simple[4], together with the common roots of

$$\sin \beta = 0, \quad \sin \sigma \beta = 0. \tag{14b}$$

However, these common roots, if they exist, do not contribute to the sums (12a) and (12b) because of the presence of $\sin\beta_m$ and $\sin\sigma\beta_m$ factors in the numerators of these expressions. Therefore the sums in (12) run only over the roots of eq. (14a).

Moreover, we can always approximate $\sigma$ as a ratio of two integers namely, $C = r/p$. In this case the roots of (14a) can be written as

$$\beta_m \to \beta_{v,n} = \beta_v + n(p\pi), \; n = 1, 2, ..., \tag{15}$$

with $\beta_v$ representing the roots of (14a) on the interval from 0 to $p\pi$. Taking into account (15) and using (14a) we can write the temperature distribution in the rods in the final form as

$$T_1(x, t) = T_{eq} -$$

$$- 2(T_{02} - T_{01}) \sum_v \frac{\sin \beta_v \cos \beta_v \sin^2 \sigma\beta_v}{\sin^2 \sigma\beta_v + \kappa\sigma \sin^2 \beta_v} \sum_{n=0}^{\infty} \frac{e^{-\frac{\chi_1 t}{L^2}(\beta_v + np\pi)^2} \cos[(\beta_v + np\pi)(x/L)]}{\beta_v + np\pi}$$

$$+ 2(T_{02} - T_{01}) \sum_v \frac{\sin^2 \beta_v \sin^2 \sigma\beta_v}{\sin^2 \sigma\beta_v + \kappa\sigma \sin^2 \beta_v} \sum_{n=0}^{\infty} \frac{e^{-\frac{\chi_1 t}{L^2}(\beta_v + np\pi)^2} \sin[(\beta_v + np\pi)(x/L)]}{\beta_v + np\pi} \tag{16a}$$

and

$$T_2(x, t) = T_{eq} +$$

$$+ 2\kappa\sigma (T_{02} - T_{01}) \sum_v \frac{\sin \sigma\beta_v \cos \sigma\beta_v \sin^2 \beta_v}{\sin^2 \sigma\beta_v + \kappa\sigma \sin^2 \beta_v} \sum_{n=0}^{\infty} \frac{e^{-\frac{\chi_2 t}{L^2}(\sigma\beta_v + nr\pi)^2} \cos[(\sigma\beta_v + nr\pi)(x/L)]}{\sigma\beta_v + nr\pi}$$

$$+ 2\kappa\sigma (T_{02} - T_{01}) \sum_v \frac{\sin^2 \sigma\beta_v \sin^2 \beta_v}{\sin^2 \sigma\beta_v + \kappa\sigma \sin^2 \beta_v} \sum_{n=0}^{\infty} \frac{e^{-\frac{\chi_2 t}{L^2}(\sigma\beta_v + nr\pi)^2} \sin[(\sigma\beta_v + nr\pi)(x/L)]}{\sigma\beta_v + nr\pi}. \tag{16b}$$

**Case 2:** Since $L_1/\sqrt{\chi_1} = L_2/\sqrt{\chi_2}$, the results (8) simplify to the following expressions,

$$\tilde{T}_1(x,s) = \frac{T_{01}}{s} + \frac{T_{02} - T_{01}}{(\kappa+1)s} \cdot \frac{\cosh\left((x+L_1)\sqrt{\frac{s}{\chi_1}}\right)}{\cosh\left(L_1\sqrt{\frac{s}{\chi_1}}\right)}, \tag{17a}$$

$$\tilde{T}_2(x,s) = \frac{T_{02}}{s} - \frac{\kappa(T_{02} - T_{01})}{(\kappa+1)s} \cdot \frac{\cosh\left((x-L_2)\sqrt{\frac{s}{\chi_2}}\right)}{\cosh\left(L_2\sqrt{\frac{s}{\chi_2}}\right)}. \tag{17b}$$

The corresponding inverse Laplace transforms can be found in Ref. 3, p. 213, for example, and are

$$T_1(x,t) = \frac{\kappa T_{01} + T_{02}}{\kappa+1} + \frac{2(T_{02} - T_{01})}{\kappa+1} \cdot \frac{1}{\pi} \sum_{n=1}^{\infty} \frac{e^{-(n-\frac{1}{2})^2 \pi^2 \chi_1 t / L_1^2}}{(n-\frac{1}{2})} \sin\left((n-\tfrac{1}{2})\frac{\pi x}{L_1}\right), \tag{18a}$$

$$T_2(x,t) = \frac{\kappa T_{01} + T_{02}}{\kappa+1} + \frac{2\kappa(T_{02} - T_{01})}{\kappa+1} \cdot \frac{1}{\pi} \sum_{n=1}^{\infty} \frac{e^{-(n-\frac{1}{2})^2 \pi^2 \chi_2 t / L_2^2}}{(n-\frac{1}{2})} \sin\left((n-\tfrac{1}{2})\frac{\pi x}{L_2}\right). \tag{18b}$$

### III. NUMERICAL RESULTS AND DISCUSSION

From eqs. (1) and (18) it follows that the interface temperature is constant in the case of semi-infinite rods and rods with their length ratio $L_1/L_2 = \sqrt{\chi_1/\chi_2}$, and is equal to the value given by eq. (2). On the other hand, the interface temperature for the rods of equal length is, for example,

$$T_1(0,t) = T_{eq} - 2(T_{02} - T_{01}) \sum_{\nu} \frac{\sin\beta_\nu \cos\beta_\nu \sin^2 \sigma\beta_\nu}{\sin^2 \sigma\beta_\nu + \sigma\kappa \sin^2 \beta_\nu} \sum_{n=0}^{\infty} \frac{e^{-\frac{\chi_1 t}{L^2}(\beta_\nu + np\pi)^2}}{\beta_\nu + np\pi}. \tag{19}$$

It is plotted in Fig. 1 as a function of $\chi_1 t/L^2$. We observe, however, that even in this case the interface temperature whose initial value is again given by eq. (2) remains approximately constant for a relatively large fraction of time $L^2/\chi$, which is the time characteristic for the rods to reach thermal equilibrium. This initial and apparently general behavior of interface temperature is understandable because for times such that $\chi t \ll L^2$ the heat flow is independent of the length of the rods. For semi-infinite rods this condition is clearly satisfied at all times and, consequently, the interface temperature is thus rigorously constant. For finite rods with arbitrary ratio of their lengths the interface temperature, of course, starts to change with increasing time and eventually approaches the equilibrium value given by eq. (10). This is illustrated in Fig. 2 showing the temperature distribution along the rods of equal lengths for various values of time. For comparison, the temperature distribution in semi-infinite rods is also shown in Fig. 3 and, for finite rods with the length ratio $L_1/L_2 = \sqrt{\chi_1/\chi_2}$, in Fig. 4.

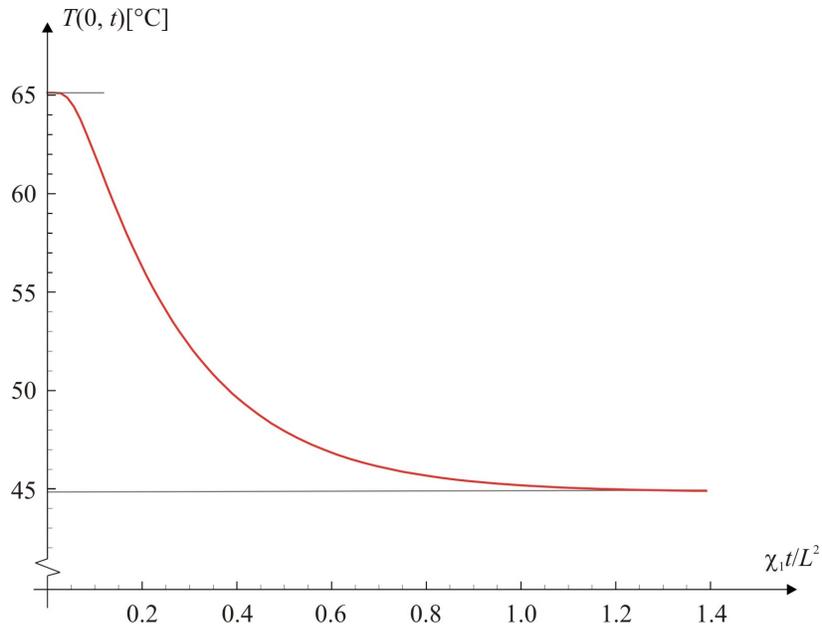

Fig. 1. Temperature at the contact for the case of a copper (initial temperature $T_{01} = 10$ °C) and aluminum rod (initial temperature $T_{02} = 100$ °C), both of length $L = 1$ m, as a function of $\chi_1 t/L^2$. The upper horizontal line is the initial contact temperature $T_1(0, t > 0) = T_2(0, t > 0) = 65.12$ °C given by Eq. (2), the lower horizontal line is the final equilibrium temperature $T_{eq} = 44.85$ °C given by Eq. (3).

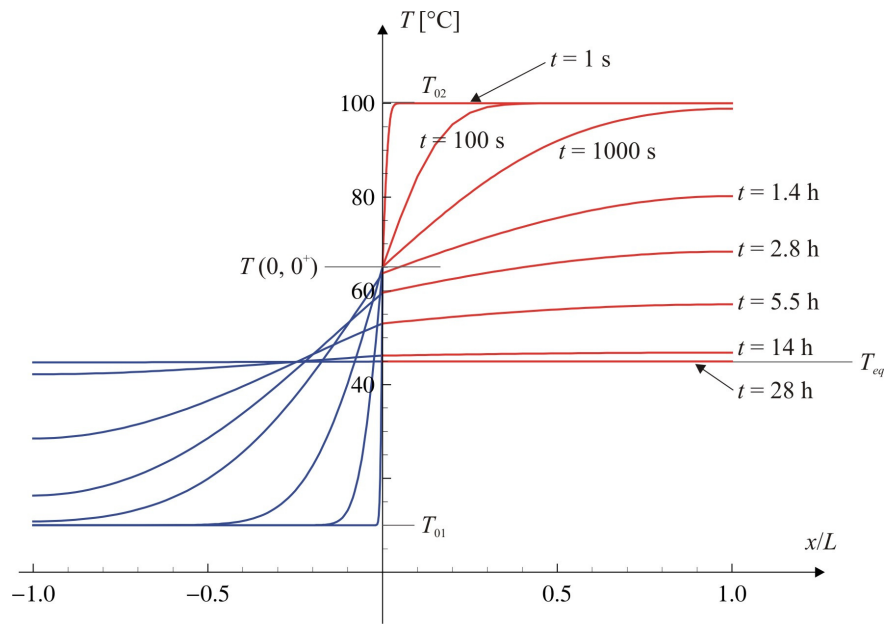

Fig. 2. Spatial distribution of temperature ($T = T(x, t)$) in a copper and an aluminum rod, both of length $L = 1$ m, in contact at $x = 0$, for various times: $t = 1$ s, 100 s, 1000 s, 5000 s, 10000 s, 20000 s, 50000 s and 100000 s. The initial temperature at the contact is $T(0, t = 0^+) = 65.12$ °C, the final (equilibrium) temperature $T_{eq} = T(0, t \to \infty) = 44.85$ °C.

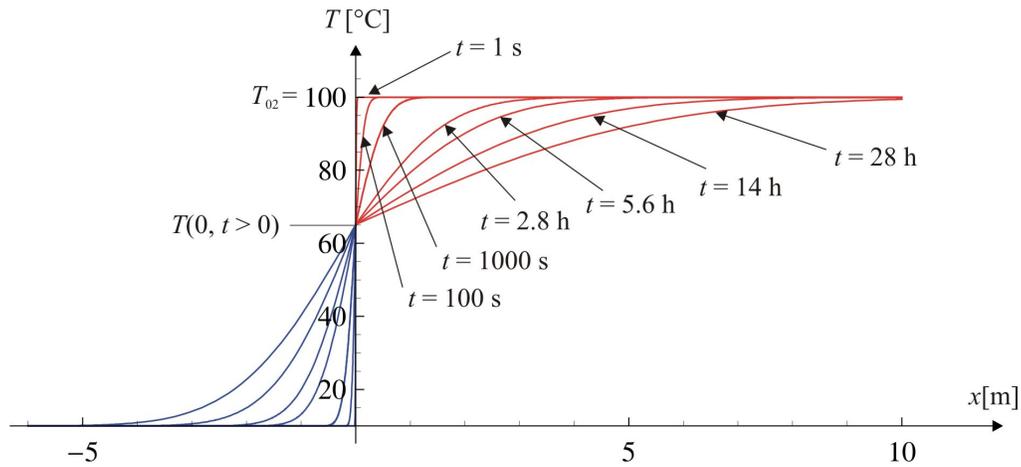

Fig. 3. Spatial distribution of temperature ($T = T(x, t)$) in a copper and aluminum rod of semi-infinite length, in contact at $x = 0$, for various times: $t = 1$ s, 100 s, 1000 s, 10000 s, 20000 s, 50000 s and 100000 s. The temperature at the contact ($T(0, t > 0)$) is 65.12 °C at all times.

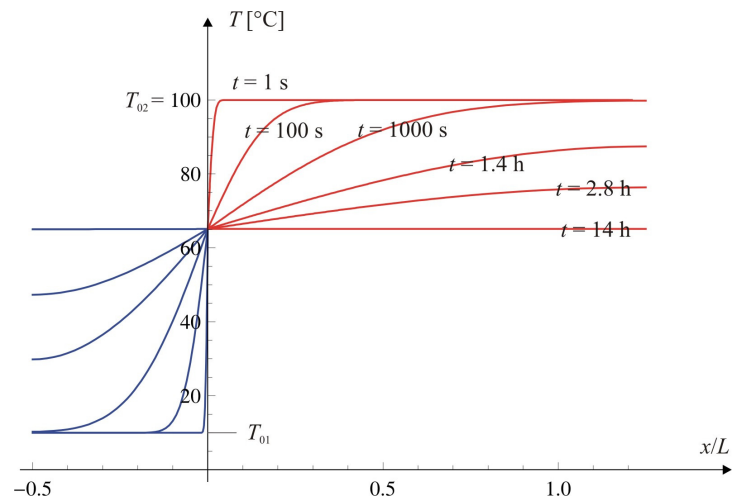

Fig. 4. Spatial distribution of temperature ($T = T(x, t)$) in a copper and aluminum rod of length ratio $L_1/L_2 = \sqrt{\chi_1 / \chi_2}$, in contact at $x = 0$, for various times: $t = 1$ s, 100 s, 1000 s, 10000 s, 20000 s, 50000 s and 100000 s. The temperature at the contact does not change and is at all times 65.12 °C which is also the final equilibrium temperature of the rods.

On the basis of these results we conclude that the interface temperature is given initially, in all cases, by the value obtained from eq. (2). Moreover, we have shown that this value corresponds to the thermal equilibrium value $T'_{eq}$ of two finite rods whose length ratio is $L_1/L_2 = \sqrt{\chi_1/\chi_2}$. Thai is,

$$T'_{eq} = \frac{L_1\rho_1 c_1 T_{01} + L_2\rho_2 c_2 T_{02}}{L_1\rho_1 c_1 + L_2\rho_2 c_2} = \frac{\left((L_1/L_2)\sqrt{\chi_2/\chi_1}\right)\kappa T_{01} + T_{02}}{\left((L_1/L_2)\sqrt{\chi_2/\chi_1}\right)\kappa + 1} = \frac{\kappa T_{01} + T_{02}}{\kappa + 1}, \tag{20}$$

in agreement with eq. (10). The condition that the interface temperature remains approximately constant is $\chi_1 t \ll L_1^2$, $\chi_2 t \ll L_2^2$. When this condition ceases to be satisfied, the interface temperature starts to change, except in the case of semi-infinite rods, where this condition is never violated, and in the particular case of finite rods with their length ratio adjusted in such a way that the equilibrium temperature $T'_{eq}$ coincides with the initial interface temperature.

The equilibrium temperature for semi-infinite rods must therefore be compared to the equilibrium temperature of two finite rods of lengths $L_1$ and $L_2$ with their ratio equal to $L_1/L_2 = \sqrt{\chi_1/\chi_2}$. The transition to the limit $L_1, L_2 \to \infty$, determined by the heat conduction process itself, is such that this ratio is kept fixed.

Thus, when we join two rods or two bodies with different temperatures, small regions on each side of the interface whose sizes are roughly of the order $\sqrt{\chi_1 t}$ and $\sqrt{\chi_2 t}$, respectively, almost immediately reach the equilibrium temperature (20). The respective portion of the body with initial temperature $T_{02} > T_{01}$ cools by the amount

$$T_{02} - T'_{eq} = \frac{\kappa}{1+\kappa}(T_{02} - T_{01}), \tag{21a}$$

while the respective portion of the colder body (with initial temperature $T_{01}$) heats up by the amount

$$T'_{eq} - T_{01} = \frac{T_{02} - T_{01}}{1+\kappa}, \tag{21b}$$

where $\kappa = \frac{k_1}{k_2}\sqrt{\frac{\chi_2}{\chi_1}} = \sqrt{\frac{k_1\rho_1 c_1}{k_2\rho_2 c_2}}$. A practical example illustrating this result is the following well-known situation. When we step with bare feet ($T_{02}$) on a cold surface ($T_{01}$) our feet will cool off by an amount given by (21a) which is the larger the larger is $\kappa$. Consequently, those surfaces for which $\kappa$, as defined above, is large will be perceived as colder than surfaces with small $\kappa$ even though their temperatures are the same.

**APPENDIX**

In the case of temperature diffusion in semi-infinite rods aligned with the $x$-axis the only quantity with dimension of length in addition, of course, to the coordinate $x$ is $\sqrt{\chi t}$. Consequently the only dimensionless quantity on which the temperature distribution may depend is $x/\sqrt{\chi t}$. Since the diffusion equation is linear and homogeneous with respect to temperature its unit can be assigned arbitrarily. If $T_0$ is some temperature characteristic to the problem at hand, we may write

$$\frac{T(x,t)}{T_0} = f\left(\frac{x}{2\sqrt{\chi t}}\right) \equiv f(\xi), \tag{A1}$$

where we have included a factor of 2 for later convenience. This transformation is originally due to Boltzmann[5]. Inserting this function into diffusion equation (4) we obtain

$$-\frac{1}{2}\xi \frac{df}{d\xi} = \frac{1}{4}\frac{d^2 f}{d\xi^2}. \tag{A2}$$

Assuming that two rods with initial temperatures $T_1$ (left rod) and $T_2$ (right rod) are joined at $x = 0$ and integrating the above equation, we write the solutions in the form

$$T(x<0,t) = A_1 + B_1 \frac{2}{\sqrt{\pi}} \int_{x/2\sqrt{\chi_1 t}}^{0} e^{-\xi^2} d\xi = A_1 + B_1 erf\left(\frac{-x}{2\sqrt{\chi_1 t}}\right), \tag{A3}$$

$$T(x>0,t) = A_2 + B_2 \frac{2}{\sqrt{\pi}} \int_{0}^{x/2\sqrt{\chi_2 t}} e^{-\xi^2} d\xi = A_2 + B_2 erf\left(\frac{x}{2\sqrt{\chi_2 t}}\right), \tag{A4}$$

where $erf(x)$ is the error function defined as

$$erf(x) = \frac{2}{\sqrt{\pi}} \int_{0}^{x} e^{-\xi^2} d\xi,$$

$$erf(\infty) = 1,$$
$$erf(-x) = -erf(x).$$

Imposing the initial and boundary conditions (5) enables us to determine the integration constants with the results

$$A_1 = \frac{\kappa T_1 + T_2}{1+\kappa}, \quad B_1 = \frac{T_1 - T_2}{1+\kappa}, \quad A_2 = A_1, \quad B_2 = -\kappa B_1.$$

Inserting these expressions into (A3) and (A4) we obtain the results quoted in the introduction.